\documentclass[final]{elsart}

\usepackage{graphics}
\usepackage{graphicx}

\pdfoutput=1

\newcommand{\FIG}[1]{#1}

\newcommand{\vv}{\mbox{\bf v}}

\newcommand{\UU}{\mbox{\bf U}}
\newcommand{\FF}{\mbox{\bf F}}
\newcommand{\BB}{\mbox{\bf B}}

\newcommand{\bfB}{\mbox{\bf B}}

\newcommand{\bfS}{\mbox{\bf S}}

\begin{document}

\begin{frontmatter}

\title{A multidimensional grid-adaptive relativistic magnetofluid code}
\author[Leuven]{B. van der Holst},
\ead{Bart.vanderHolst@wis.kuleuven.be}
\author[Leuven,FOM,Utrecht]{R. Keppens}, and
\author[FOM]{Z. Meliani}

\address[Leuven]{Centre for Plasma-Astrophysics, K.U.Leuven, Celestijnenlaan
200B, 3001 Heverlee, Belgium}
\address[FOM]{FOM-Institute for Plasma Physics Rijnhuizen, P.O. Box 1207, 3430
BE Nieuwegein, The Netherlands}
\address[Utrecht]{Astronomical Institute, Utrecht University, P.O. Box 80000,
3508 TA Utrecht, The Netherlands}

\begin{abstract}

A robust second order, shock-capturing numerical scheme for multi-dimensional
special relativistic magnetohydrodynamics on computational domains with adaptive
mesh refinement is presented. The base solver is a total variation diminishing
Lax-Friedrichs scheme in a finite volume setting and is combined with a
diffusive approach for controlling magnetic monopole errors.
The consistency between the primitive and conservative variables is ensured
at all limited reconstructions and the spatial part of the four velocity
is used as a primitive variable. Demonstrative relativistic examples are
shown to validate the implementation. We recover known exact solutions to
relativistic MHD Riemann problems, and simulate the shock-dominated long term
evolution of Lorentz factor 7 vortical flows distorting magnetic island chains.

\end{abstract}

\end{frontmatter}

\section{Introduction}

Relativistic flows of magnetized plasmas are observed in a wide variety of
astrophysical objects. At galactic scales, accretion of matter around
black holes in Active Galactic Nuclei (AGN) in addition results
in well collimated jets emitted along the rotation axis. To explain their observed superluminal
plasma motion at the parsec length scales, flows with Lorentz factors
of more than 10 are needed. The observed
synchrotron emission indicates that these jets are pervaded by magnetic fields.
Even more powerful processes are at play in the highly relativistic
blast waves associated with Gamma Ray Bursts (GRBs).
Here, the plasma flows easily reach Lorentz factors of 100 or even higher.
The morphology and time evolution of these and other astrophysical objects
often involve strong shocks and complex magnetic field topologies. To compute this
kind of challenging relativistic plasma dynamics, the combination of Adaptive Mesh Refinement (AMR)
with a robust, shock-capturing numerical method is therefore indispensible.

With growing interest in relativistic astrophysical phenomena, 
various efforts are ongoing to develop numerical
special relativistic hydrodynamic and magnetohydrodynamic codes. Significant progress
was achieved in the last decade with the development of conservative shock-capturing schemes, 
which use either exact Riemann solvers 
or approximate Riemann solvers, or more robust central type schemes, in relativistic hydrodynamics~\cite{Eulderink&Melemma94, Fontetal94, Donatetal, Aloyetal99, delzanna02, Marti&Muller03, Mignone&Bodo05} (for a contemporary review see \cite{Marti&Muller03}) 
and in relativistic magnetohydrodynamics 
\cite{komis, balsara, koldoba, delzanna, jeroen, leis, Mignone&Bodo06, mignone2007, hllc07}. 
The study of relativistic hydrodynamic fluids currently starts to benefit also from using spatial and temporal adaptive techniques, or Adaptive Mesh Refinement (AMR) \cite{Zhang&MacFadyen06, Melianiet07, Wangetal07}.
To date, AMR was incorporated only in some works, both in special relativistic 
magnetohydrodynamics \cite{jeroen} and in general relativistic magnetohydrodynamics \cite{Anninosetal05,Andersonetal06,giacom2}.
At the same time, various authors, including \cite{Melianiet07}, have shown that AMR is imperative to simulate extreme astrophysical phenomena, such as those encountered in GRBs. 

In order to handle GRB and other extreme relativistic flow regimes, the employed solver needs to be very robust under a wide 
variety of plasma conditions. Therefore, we decided to use a conservative discretization which does not fully exploit the 
detailed solution knowledge of the Riemann problem at each cell interface,
defined by two constant states in contact. This contrasts with a true Godunov scheme, which would compute the flux across cell
interfaces from the exact solution to the local Riemann problem. 
While fairly recently~\cite{giacom}, an exact Riemann problem solver for the RMHD equations became available, the nonlinear
iteration involved would make it a fairly computationally costly
ingredient for a multidimensional code. As pointed out further, we already face a similar Newton-Raphson procedure
to deduce primitive from conservative variables in every grid point. In fact, we follow the trend away from using exact or 
approximate Riemann problem solvers in most codes in use to date. In the works cited above, Komissarov~\cite{komis} already 
used a
linearized (approximate) Riemann solver, and even fell back on an HLLE variant~\cite{hll83, einf88} which only uses the fastest
wave speeds. Koldoba et al.~\cite{koldoba} and Balsara~\cite{balsara} also employ an approximate, linearized Riemann solver and presented 
numerical solutions for stringent 1D test problems. All other works mentioned~\cite{delzanna, jeroen, leis, Mignone&Bodo06, hllc07} use central-type schemes, of either the local
Lax-Friedrichs, HLL, or HLLC variant, suitably adopted to the relativistic MHD regime. We will use the simplest of these
in our implementation, since in combination with automated grid adaptivity, we rely on convergence and robustness of the underlying
discretization, while accuracy is most easily gained by raising the effective resolution employed.

The macroscopic dynamics of astrophysical objects with relativistic plasma flows are
governed by the conservation of particle and momentum-energy, along with the Maxwell
equations to account for the involvement of the magnetic field. In Sect.~\ref{sec-eq}, we
recast the RMHD equations in conservation form. Details on the implementation
of the high-resolution shock-capturing scheme and the adaptive mesh refinement
in our code can be found in Sects.~\ref{sec-scheme}-\ref{sec-amr}. The code is validated against several known
1D analytical solutions in Sect.~\ref{sec-tests} and two multi-dimensional relativistic
flow simulations are performed. In Sect.~\ref{sec-summ}, a summary of the paper containing
our main conclusions and an outlook for future applications of the code is given.

\section{Governing Equations}\label{sec-eq}

We give a brief outline of the derivation of the relativistic magnetohydrodynamic (RMHD)
equations and restrict the analysis to special relativity only. For a more
extensive account we refer to~\cite{komis}. Along the way, we point out minor differences in the choice of conservative variables, as well as
a possibility to exploit an entropy related variable. The dynamics follow from the
Maxwell equations and the conservation of particle number density and energy-stress.
We will fix a Lorentzian reference frame to perform the computations.
Due to Lorentz contraction the fluid volume elements are contracted so
that the particle number density is multiplied by the Lorentz factor
$\Gamma=(1-{\bf v}^2)^{-1/2}$. Here, ${\bf v}=(v_x, v_y, v_z)$ is the spatial
velocity three-vector. We measure in the often used scaling for which the speed of light is unity $c=1$.
The particle number conservation equation can generally be written as
\begin{equation}
\partial_\alpha(\rho u^\alpha)=0,
\label{eq-density}
\end{equation}
where $\rho$ is the proper mass density, i.e., the density in that Lorentz frame in which
the fluid is locally at rest, $u^\alpha=\Gamma(1,{\bf v})$ is the four velocity
that has to satisfy the constraint $u_\alpha u^\alpha=-1$, and Greek indices denote
four vectors. We will use Latin indices to denote three-vectors. The Einstein summation convention
for repeated indices is assumed.

The antisymmetric dual field tensor $\mathcal{F}^{\alpha\beta}$ can be defined in terms of the
magnetic field ${\bf B}$ and electric field ${\bf E}$ as measured in the Lorentzian reference frame by
$\mathcal{F}^{\alpha\beta}=-\mathcal{F}^{\beta\alpha}$, $\mathcal{F}^{0i}=B^i$,
$\mathcal{F}^{ij}=-\varepsilon^{ijk}E_{k}$, where $\varepsilon^{ijk}$ is the Levi-Cevita
symbol. The homogeneous Maxwell equations can be expressed in terms of this dual tensor as
\begin{equation}
 \partial_\beta\mathcal{F}^{\alpha\beta}=0.
\end{equation}
By subsequently assuming the fluid to be perfectly conducting, so that the comoving electric field vanishes or ${\bf E}+{\bf v}\times{\bf B}=0$,
and by introducing the four-vector
\begin{equation}
 b^{\alpha}=\mathcal{F}^{\alpha\beta}u_\beta=(\Gamma{\bf v}\cdot{\bf B},
            {\bf B}/\Gamma+{\Gamma\bf v}\cdot{\bf B}{\bf v}),
\end{equation}
the induction equation and $\nabla\cdot{\bf B}=0$ transform to
\begin{equation}
 \partial_\alpha(u^\alpha b^\beta - b^\alpha u^\beta)=0.
\label{eq-induction}
\end{equation}

Finally, conservation of energy and momentum follow from
\begin{equation}
 \partial_\beta(T_{\rm fl}^{\alpha\beta}+T_{\rm em}^{\alpha\beta})=0.
\label{eq-stressenergy}
\end{equation}
Here
\begin{eqnarray}
 T_{\rm fl}^{\alpha\beta} &=& \rho hu^{\alpha}u^{\beta}+p\eta^{\alpha\beta},\\
 T_{\rm em}^{\alpha\beta} &=& |b|^2(u^{\alpha}u^{\beta}
                              +\frac{1}{2}\eta^{\alpha\beta})-b^{\alpha}b^{\beta},
\end{eqnarray}
are the fluid and electromagnetic energy-stress tensor, $\eta^{\alpha\beta}={\rm diag}(-1,1,1,1)$
is the metric tensor in flat Minkowski space-time, and $h$ is the relativistic specific enthalpy
of the fluid in the local comoving frame. In the derivation of the electromagnetic tensor,
use has been made of the Maxwell equations.

To make the equations more useful for the numerical approach based on temporally advancing conservation laws, we split the equations
for the mass density (\ref{eq-density}), momentum and energy (\ref{eq-stressenergy}),
and the magnetic field (\ref{eq-induction}) in the space and time coordinates:
\begin{eqnarray}
 & & \partial_t(\Gamma\rho)+\partial_i(\Gamma\rho v^i)=0, \label{eq-cons1}\\
 & & \partial_t(\Gamma^2\rho h_{\rm tot}v^j-b^0b^j)+\partial_i(\Gamma^2\rho h_{\rm tot}v^iv^j
     +p_{\rm tot}\delta^{ij}-b^ib^j)=0, \\
 & & \partial_t(\Gamma^2\rho h_{\rm tot}-p_{\rm tot}-(b^0)^2)+\partial_i(\Gamma^2\rho h_{\rm tot}v^i
     -b^0b^i)=0, \\
 & & \partial_t B^j+\partial_i(v^iB^j-B^iv^j)=0, \qquad \partial_i B^i=0. \label{eq-cons8}
\end{eqnarray}
In these equations, we have introduced the relativistic magnetic and total pressure
\begin{eqnarray}
p_{\rm mag} &=& \frac{1}{2}(\frac{B^2}{\Gamma^2}+({\bf v}\cdot{\bf B})^2), \\
p_{\rm tot} &=& p+p_{\rm mag},
\end{eqnarray}
where $p$ is the thermal pressure in the local comoving frame. The total relativistic
specific enthalpy is given by
\begin{equation}
h_{\rm tot} = h+\frac{2p_{\rm mag}}{\rho}.
\end{equation}
The equations (\ref{eq-cons1})--(\ref{eq-cons8}) are explicitly in conservation form. We actually combine
the first and third equation as written above to arrive at
\begin{equation}
  \partial_t\UU + \nabla \cdot\FF(\UU) = 0,
\label{eq-conseq}
\end{equation}
where
\begin{eqnarray}
\label{eq-consvars}
 \UU & = & \left( D, \bfS, E, \bfB\right)^T \nonumber\\
  & = & \left( \Gamma \rho, (\xi + B^2)\vv -(\vv\cdot\BB)\BB, \right. \nonumber\\
 & & \left. \xi + \frac{B^2}{2}+\frac{1}{2}(v^2B^2-(\vv\cdot\BB)^2)-p-D,
\bfB\right)^T , \\
\label{eq-consfluxes}
 \FF & = & \left( D \vv ,
(\xi+B^2)\vv\vv-\frac{\BB\BB}{\Gamma^2}-(\vv\cdot\BB)(\BB\vv+\vv\BB)+Ip_{\rm
tot}, \right. \nonumber \\
 & & \left. E \vv+p_{\rm tot}\vv-(\vv\cdot\BB)\BB, \vv\BB-\BB\vv \right)^T,
\end{eqnarray}
Here, $\Gamma D=\Gamma^2\rho$ is the mass density in the inertial reference frame
and $\xi=\Gamma^2\rho h$ is a measure for the enthalpy. The variable $E$ has a contribution from the
rest mass density subtracted from the total energy density as written for the laboratory frame, and represents a small
difference between our variables set and the one used in most implementations~\cite{delzanna, komis, balsara}. This makes the current set of
variables reduce to the usual set employed in non-relativistic MHD computations in the limit of $\Gamma \rightarrow 1$. This
also helps to avoid potential numerical problems with negative pressures, in cases when the rest mass contribution is a dominant term in the energy density.
In the case of an ideal
equation of state with a constant polytropic index $\gamma$, the variable $\xi$ becomes
\begin{equation}
\xi = \Gamma^2\left(\rho+\frac{\gamma p}{\gamma-1}\right).
\label{eq-xidef}
\end{equation}
The quantity $\xi$ (and $\rho h$) reduces to the non-relativistic enthalpy, except for
the inclusion of the rest mass contribution.

It is noted that instead of choosing the energy density $E$ as a conserved variable,
we can also switch to $DS$, where $S=p/\rho^\gamma$ is the specific entropy. The energy
equation is then replaced by
\begin{equation}
\partial_t(DS)+\nabla\cdot(DS{\bf v})=0,
\end{equation}
expressing the conservation of entropy for ideal RMHD. This conservation law
also follows from $u_\alpha\partial_\beta T_{\rm em}^{\alpha\beta}=
b_\alpha\partial_\beta\mathcal{F}^{\alpha\beta}=0$, so that
$u_\alpha\partial_\beta T_{\rm fl}^{\alpha\beta}=u_\alpha\partial_\beta T^{\alpha\beta}=0$
leading after some algebra to the conservation of entropy.
The energy equation has the distinct disadvantage that numerical calculations with small
kinetic pressure compared to magnetic pressure can, depending on the $\nabla\cdot{\bf B}$
strategy in use, easily
result in negative pressure. This problem could be circumvented by switching to the entropy equation.
The use of this entropy conservation law was also mentioned in~\cite{koldoba}, where it is advised for use in applications
without shocks. Even on a fixed grid, a strategy for using this equation locally instead of the energy density law
would need to be provided, explaining when positivity is preferred above strict conservation. Examples of such switching strategies can be found in challenging cosmological (non-relativistic) hydro codes, such as in~\cite{Ryu93, bryan}. In the shock-dominated simulations described below, we as yet did not need to use such a switching strategy. However, the grid adaptive
code employed here allows to use a related idea, but only at the time of
restriction and prolongation actions. Before every restriction/prolongation, we have the option to locally convert to the set of
conserved variables $(D,\bfS, DS, \bfB)$, keeping both conservation and positivity garantueed in these operations. After the coarsening or refining action, we then revert to the
original conservative set employed in the time integration routines. This makes this strategy particular to AMR
computations which can involve strong shocks, such as some of the tests presented below.
While we did use this in some of our non-relativistic MHD computations so far, the tests mentioned below were
performed without the need to invoke this entropy strategy.

\section{Relativistic Shock Capturing Scheme}\label{sec-scheme}

\subsection{Switching from Conservative to Primitive Variables}

To advance the set of conservation equations (\ref{eq-conseq}--\ref{eq-consfluxes}) in time,
we need to calculate the fluxes. The latter are obtained from the primitive variables
$(\rho, {\bf v}, p, {\bf B})$. While the determination of the primitive variables for
non-relativistic MHD is a straightforward algebraic manipulation, the transformation
of the variables for the relativistic MHD equations needs a root finding algorithm.
The analysis is facilitated by the introduction of the auxiliary variables $(\Gamma,\xi)$,
which are to be determined from the conservative variables $(D,{\bf S}, E, {\bf B})$ only.
Once the auxiliary variables are known, the construction of the primitive variables will be
straightforward.

We follow the method given in \cite{jeroen} to determine $\xi$. From the definition of the conservative
variables (\ref{eq-consvars}) and $(\Gamma,\xi)$, we obtain the primitive
variables
\begin{equation}
{\bf v}=\frac{{\bf S}+\xi^{-1}({\bf S}\cdot{\bf B}){\bf B}}{\xi+B^2},
\label{eq-velocity}
\end{equation}
and
\begin{equation}
\rho=D/\Gamma, \qquad p=\frac{\gamma-1}{\gamma}\frac{(\xi-\Gamma D)}{\Gamma^2}.
\end{equation}
Using the definition of $E$ in Eq.\ (\ref{eq-consvars}) and the expression of $\xi$ (\ref{eq-xidef}),
it follows that $\xi$ is the root of 
\begin{equation}
f(\xi)=\xi-\frac{\gamma-1}{\gamma}\frac{(\xi-\Gamma D)}{\Gamma^2}-E-D+B^2-\frac{1}{2}
   \left[ \frac{B^2}{\Gamma^2}+\frac{({\bf S}\cdot{\bf B})^2}{\xi^2} \right]=0,
\label{eq-rootfinding}
\end{equation}
where the second term is the kinetic pressure $p$.
The still unknown Lorentz factor $\Gamma=\Gamma({\bf S},{\bf B};\xi)$ contains once more the variable $\xi$
as follows from Eq.\ (\ref{eq-velocity}):
\begin{equation}
\frac{1}{\Gamma^2}=1-v^2=1-\frac{ {\left| {\bf S}+\xi^{-1}({\bf S}\cdot{\bf B}){\bf B} \right|}^2 }{(\xi+B^2)^2}.
\label{eq-augmentedvar}
\end{equation}
In the root finding algorithm, $\xi$ has to be found as a zero of Eq.\ (\ref{eq-rootfinding}) with the help
of the augmented equation (\ref{eq-augmentedvar}).

The root $\xi$ is found by means of a Newton-Raphson method. In our implementation, the brackets for
the Newton-Raphson method are found as follows: We constrain the pressure by a given lower bound $p_\epsilon$,
so that the values of $\xi$ are restricted from below by
\begin{equation}
\xi\equiv\Gamma^2(\rho + \frac{\gamma}{\gamma-1}p)> D+\frac{\gamma}{\gamma-1}p_\epsilon\equiv\xi_1,
\end{equation}
since $\Gamma\geq1$. The velocity that follows from this bound on $\xi_1$ is constrained to a
value below the speed of light, that means, by using Eq.\ (\ref{eq-augmentedvar})
\begin{equation}
v^2(\xi_1)\leq(1-dv_\epsilon)^2,
\end{equation}
where $dv_\epsilon$ represents a given threshold. If $\xi_1$ does not satisfy this condition, then
find a new $\xi_1$ via a Newton-Raphson procedure from $v^2(\xi)=(1-dv_\epsilon)^2$.
The obtained $\xi_1$ is our lower bound. As a maximal bound for the bracketing we choose
$\xi_2=E+D+p_\epsilon$. Next, we check the sign of $f(\xi_1)$ and $f(\xi_2)$. If they are the same,
then the brackets are wrong, and a new guess for the brackets is found by successively replacing
$\xi_1$ by $\xi_2$ and $\xi_2$ by $2\xi_2$. We follow this procedure till we have
correct brackets. Finally, we apply the Newton-Raphson method to find $\xi$ and $\Gamma$,
followed by a consistency check to ensure that $v<1$ and $p\geq p_\epsilon$.

\subsection{Determining the Characteristic Speeds}

The shock-capturing scheme used in this work needs the (maximal) characteristic speeds of a given state of a fluid
element. Since the equations are formulated in the inertial reference frame, we need to determine
the characteristic wave speeds in this frame.

For each spatial dimension $i$ the RMHD equations (\ref{eq-cons1})--(\ref{eq-cons8})
yields seven characteristic speeds. There is one entropy wave speed
$\lambda_{\rm E}$ corresponding to the passive advection of entropy disturbances in ideal RMHD.
The other characteristic wave speeds, however, come in pairs, namely, two slow magneto-acoustic waves
speed $\lambda_{\rm S}^\pm$, two Alfv\'en wave speeds $\lambda_{\rm A}^\pm$, and two fast
magneto-acoustic wave speeds $\lambda_{\rm F}^\pm$. The characteristic speeds are ordered according to
the sequence of inequalities
\begin{equation}
\lambda_{\rm F}^- \leq \lambda_{\rm A}^- \leq \lambda_{\rm S}^- \leq \lambda_{\rm E} \leq \lambda_{\rm S}^+
\leq \lambda_{\rm A}^+ \leq \lambda_{\rm F}^+,
\end{equation}
similar to the non-relativistic MHD equations. The entropy wave is just propagating with the fluid velocity
$\lambda_{\rm E}=v_i$. The Alfv\'en disturbances propagate at speeds
\begin{equation}
\lambda_{\rm A}^\pm=v_i \pm \frac{1}{\Gamma^2}\frac{B_i}{\sqrt{\rho h_{\rm tot}}\pm({\bf v}\cdot{\bf B})},
\end{equation}
which includes relativistic corrections. The magneto-acoustic characteristic speeds follow from
the quartic polynomial
\begin{eqnarray}
& & \rho h (\frac{1}{c_{\rm s}^2}-1)\Gamma^4 (\lambda-v_i)^4 -(1-\lambda^2) \nonumber \\
& & \qquad \left\{ \Gamma^2(\rho h+\frac{2p_{\rm mag}}{c_{\rm s}^2})(\lambda-v_i)^2
   -{\left[ \Gamma({\bf v}\cdot{\bf B}) (\lambda-v_i)-\frac{B_i}{\Gamma} \right]}^2 \right\}=0,
\end{eqnarray}
and involves the relativistic hydrodynamic speed of sound $c_{\rm s}=\sqrt{\gamma p/(\rho h)}$.
These magneto-acoustic wave speeds can be found algebraically, but the formulae are not numerically efficient and
often susceptible to round-off errors. Instead we use Laguerre's method to find the four
magneto-acoustic roots in the actual implementation.
All roots are bound by the speed of light, so that they must lie in the interval
$\left] -1,1 \right[$. The roots can be located close to each other, especially when they are of
the order unity. Making the transformation $\mu=1/(1-\lambda)$, we obtain a quartic polynomial
for $\mu$. The roots are now better separated and are in the interval $\left] 0.5,\infty \right[$.
If the requested accuracy is not obtained via Laguerre's method, a 
root polishing is performed by the Newton-Raphson method.

The above mentioned scheme for finding characteristic speeds can possibly be improved if the magneto-acoustic
roots are almost degenerate. In that case root polishing by the Newton-Raphson procedure can
unintentionally make the roots degenerate. It is then preferable to switch to Maehly's procedure (see \cite{press}).
Another improvement can be made by using the transformation $\mu=1/(1-\lambda+v_i)$ to reduce
the possibility of excessively large roots. Another option which we implemented in our code is to
compute the zeros in terms $\mu=\Gamma(\lambda-v_i)$, which is somewhat better behaved for very
large Lorentz factor flows. 

\subsection{Total Variation Diminishing Lax-Friedrichs Scheme}

In the present paper, we employ the Total Variation Diminishing Lax-Fiedrichs (TVDLF) scheme~\cite{tothod} for 
relativistic MHD applications. Temporal second order accuracy is achieved by the Hancock predictor step
\begin{equation}
U_i^{n+1/2}=U_i^n-\frac{1}{2}\frac{\Delta t}{\Delta x} \left[ F(U_i^n+\frac{1}{2}\overline{\Delta U}^n_{i})
                                                             -F(U_i^n-\frac{1}{2}\overline{\Delta U}^n_{i}) \right].
\label{Hancock}
\end{equation}
Here, $\overline{\Delta U}$ denote the cell-center to cell-face limited slope used in
the TVDLF scheme.
In our work we will mostly use the rather diffusive, but stable `minmod' limiter
\begin{eqnarray}
\overline{\Delta U}_{i} &= {\rm sgn}(U_{i}-U_{i-1})& \max\left[0,\min\left\{\mid U_{i}-U_{i-1}\mid, \right. \right. \nonumber \\
& & \left. \left. (U_{i+1}-U_{i}){\rm sgn}(U_{i}-U_{i-1})\right\}\right] .
\label{minmod}
\end{eqnarray}
In the full correction step, the numerical fluxes are
\begin{eqnarray}
f^{n+1/2}_{i+\frac{1}{2}} & = & \frac{1}{2}
\left\{F(U^{L}_{i+\frac{1}{2}})+F(U^{R}_{i+\frac{1}{2}})
\right. \nonumber \\
& & \left.
-\mid c_{\rm max}(\frac{U^{L}_{i+\frac{1}{2}}+U^{R}_{i+\frac{1}{2}}}{2})\mid
\left[U^{R}_{i+\frac{1}{2}}-U^{L}_{i+\frac{1}{2}}\right]\right\},
\label{tvdlf}
\end{eqnarray}
where the left and right states are
\begin{eqnarray}
 U^{L}_{i+\frac{1}{2}} & = & U^{n+1/2}_{i}+\overline{\Delta U}^{n+1/2}_{i}/2, \nonumber \\
U^{R}_{i+\frac{1}{2}} & = & U^{n+1/2}_{i+1}-\overline{\Delta U}^{n+1/2}_{i+1}/2,
\end{eqnarray}
respectively. This TVDLF scheme does not use a Riemann solver. The only information needed is the fastest characteristic
wave speed $c_{\rm max}=\max(\left| \lambda_{\rm F}^- \right|,\left| \lambda_{\rm F}^+ \right|)$.

In this second order scheme, some improvement is obtained if the limited slopes are calculated via
the primitive variables. We have best experience  with $(\rho,\Gamma {\bf v}, p, {\bf B})$.
Note the inclusion of the Lorentz factor $\Gamma$ in the fluid velocity.
We also experimented with HLL and HLLC solvers (see e.g.~\cite{Mignone&Bodo06}), which use more information of the
Riemann fan at cell interfaces, but leave their application outside the scope of this paper. It
is our impression that the use of grid adaptivity makes the difference between these base 
solvers of secondary importance.

\subsection{Parabolic magnetic monopole treatment}\label{sec-divb}

In our multidimensional RMHD applications, we handle the $\nabla\cdot{\bf B}=0$ constraint by adding
a diffusive source term proportional to $\nabla\nabla\cdot{\bf B}$ to the induction equation.
The diffusion coefficient is determined by setting the maximum allowed diffusion time step equal to the
CFL time step. On a cartesian mesh, we obtain the source term update
\begin{eqnarray}
{\bf B} \mapsto {\bf B}+C_{\rm d}\left(\frac{1}{\Delta x^2}+\frac{1}{\Delta
y^2}+\frac{1}{\Delta z^2}\right)^{-1}\nabla\nabla\cdot\BB,
\end{eqnarray}
where $0\leq C_{\rm d} \leq 2$ for stability reasons. For most applications, a value $C_{\rm d}=1$ is advised. This is similar to the strategy used for non-relativistic MHD on curvilinear grids
in~\cite{vanderholst}.

This type of treatment for the $\nabla\cdot{\bf B}=0$ can be regarded as the parabolic variant of the hyperbolic/parabolic
treatment discussed for ideal, non-relativistic MHD in Dedner et al.~\cite{dedner}. Our source treatment redistributes potential local
monopole errors over a wider area than where they would normally concentrate. It should be noted that this is
more meant as a means to stabilize the overall numerical scheme and to avoid potential numerical instabilities, than as a manner to annihilate
discrete monopole contributions alltogether. The latter is not needed in any numerical integration, where
we will always have truncation errors in the magnetic field components, even if we numerically
ensure a kind of minimal $\nabla\cdot{\bf B}$. The hyperbolic cleaning from~\cite{dedner} in addition advects these locally
occuring numerical monopole errors, while damping them in a similar fashion. In case of shock-dominated problems (like those
presented further on), discrete monopole errors are continuously arising at shock locations, so the damping by itself is
in our opinion the most important part of the error control strategy.  The same idea was first suggested in electromagnetic PIC codes, by Marder~\cite{marder}, and is now also routinely used~\cite{nimrod} in the only 3D global tokamak plasma simulations feasible
to date, performed by the NIMROD consortium. 
One scheme for resistive relativistic MHD has recently been presented in~\cite{komis2}, where the fluxes are handled using an HLL scheme instead of our TVDLF method, while the divergence treatment there uses the Generalized Lagrange Multiplier approach from Munz et al.~\cite{munz}. This in essence is the hyperbolic variant from~\cite{dedner}, and our parabolic treatment
can be seen as belonging to the same family of monopole treatments.
In~\cite{amrvac}, a comparison of the parabolic treatment against other popular
source term strategies (Powell source terms as e.g. described in~\cite{powel99}, or only modifying the induction
equation as suggested by Janhunen~\cite{janhunen}) was performed for various multidimensional, non-relativistic MHD problems using AMR. 
The most extensive comparison of magnetic divergence control in multi-dimensional, non-relativistic MHD scenarios for fixed
grids
can be found in T\'oth~\cite{toth}. In particular, the Powell source term treatment was then tested against a variety of
constrained transport implementations, which insist on ensuring $\nabla\cdot{\bf B}=0$ to machine precision in one particular
pre-chosen discretization only, as well as against an elliptic cleaning scheme. The latter strategy projects the obtained $\bfB^*$ to a divergence free magnetic field $\bfB=\bfB^*-\nabla \phi$, and involves the solution of a Poisson equation for $\phi$. Noteworthy is that it was also found in~\cite{toth} that one does not need to enforce its solution to machine precision either.
All of these schemes were found to yield
acceptable simulation results on the nine tests verified there. Some of these tests were revisited in the grid
adaptive computations presented in~\cite{amrvac}, where only different source term treatments were compared. Those
that specifically only modify the induction equation are readily incorporated in the relativistic MHD regime,
and they do not violate the conservation of the other than magnetic variables. In our multidimensional tests below,
we quantify the remaining errors in $\nabla\cdot{\bf B}$ for illustration purposes only: the fact thay they remain
bounded at all times is the most important observation to be checked there.

\section{Adaptive Mesh Refinement}\label{sec-amr}

There are two AMR versions implemented in our AMRVAC code, namely a modified version of the original patch approach
of Berger~\cite{berger1} and a hybrid block-based~\cite{vanderholst} strategy. We will give a brief outline of these methods and refer to
the literature for details.

Once a procedure is given for detecting cells that are needed for resolving flow features, the AMR must arrange
these cells in a hierarchy of properly nested grids. In the original patch-based scheme of Berger, each cell flagged
for refinement is surrounded by a buffer-zone of a user given size to ensure that discontinuities and other
regions that need high resolution do not propagate to coarser cells. This collection of flagged cells
are then changed so that it satisfies the proper nesting: each level $l$ cell is in between level $l+1$ and $l-1$.
These cells are subsequently stored into a hierarchy of properly nested grids (patches). The patches
are then bisected till a given efficiency is reached. This efficiency is expressed as the ratio of the flagged to the
total amount of cells within a patch. Finally, a patch merging process is called to reduce the computational cost
of too many small grids. 
In the AMRVAC code, see \cite{amrvac}, the overlap of patches on the same AMR level is avoided,
while a minimal efficiency is enforced.

In the same AMRVAC code, yet another AMR scheme is implemented. This so-called hybrid block-AMR method~\cite{vanderholst}
uses, like block-based AMR techniques, an equal number of grid cells for each grid in the entire
grid hierarchy. The basic structure is an octree (in three dimensions). However, if a block is flagged
for refinement, this scheme relaxes on the standard approach where a block triggers in a $D$-dimensional
calculation $2^D$ new sub-blocks (children). This consequently introduces incomplete block families
in the grid hierarchy. Therefore, the hybrid method approaches the optimal fit of the grid
structure in the patch scheme. However, due to fixing the number of cells per grid, the good
cache performance of the common tree block-based approach is fully exploited.

The procedure to identify which cells are to be triggered for refinement relies on a type of Richardson
extrapolation. The error estimator implemented in AMRVAC is a variant of the procedure given in \cite{berger}. Given the solution vector $U_l^{n-1}$ and $U_l^n$ on level
$l$ and with a time difference $\Delta t_{l}^{n-1}$, the error estimator will first determine
$U_{l-1}^{n+1}$ in the following two ways:
\begin{itemize}
\item Coarsen $U_l^{n-1}$ and then advance to time $t=t_{l}^{n-1}+2\Delta t_{l}^{n-1}$.
\item Advance $U_l^n$ to time $t=t_{l}^{n-1}+2\Delta t_{l}^{n-1}$ and then coarsen.
\end{itemize}
The resulting solution vectors are compared and, based on a certain selection of the conservative variables
with mutual weighting factors, cells will be flagged in an automated fashion for refinement if a given
tolerance is exceeded. In addition, user-enforced refinement is possible and the auxiliary variables,
like the Lorentz factor, can be exploited in the error estimator.

Essential in the AMR strategy is the restoring of the global conservation across the entire grid
hierarchy. This amounts to fixing the fluxes of coarse cells with the fluxes of the neighboring fine
cells. Moreover, the prolongation, restriction, and/or temporal interpolation between different grid levels
need to be performed on the conservative variables. To avoid the possible introduction of negative pressure,
especially in the vicinity of small ratio of kinetic to magnetic pressure, we have implemented in
AMRVAC the option to use primitive variables during the regridding process. Another option, as pointed out earlier, would be
to switch to the entropy $DS$ instead of energy during the regridding process. Then
the AMR scheme is still conservative, but avoids the introduction of negative pressure.

\section{Numerical results}\label{sec-tests}
\subsection{Riemann problems}

\begin{figure}
\FIG{
\begin{center}
{{\resizebox{0.48\textwidth}{!}{\includegraphics{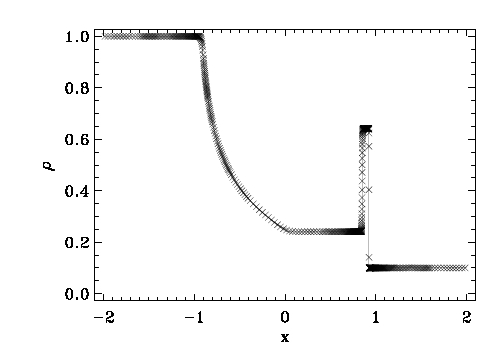}}}}
{{\resizebox{0.48\textwidth}{!}{\includegraphics{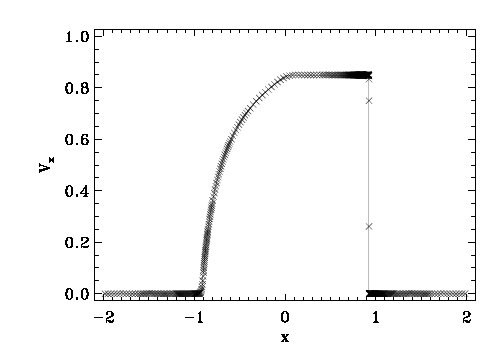}}}}
{{\resizebox{0.48\textwidth}{!}{\includegraphics{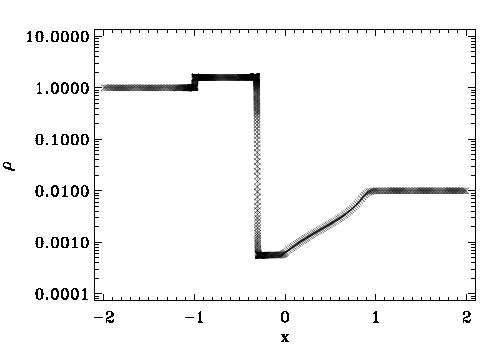}}}}
{{\resizebox{0.48\textwidth}{!}{\includegraphics{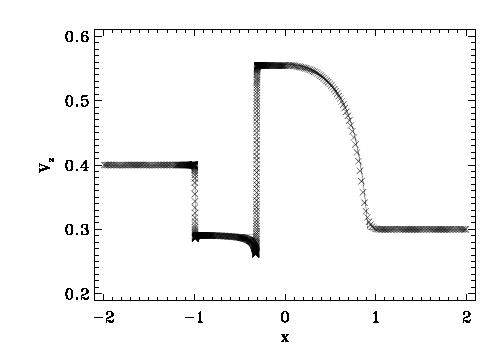}}}}
{{\resizebox{0.48\textwidth}{!}{\includegraphics{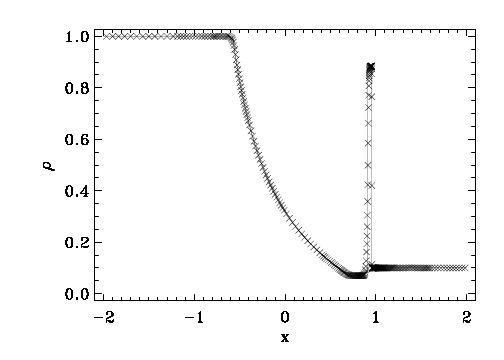}}}}
{{\resizebox{0.48\textwidth}{!}{\includegraphics{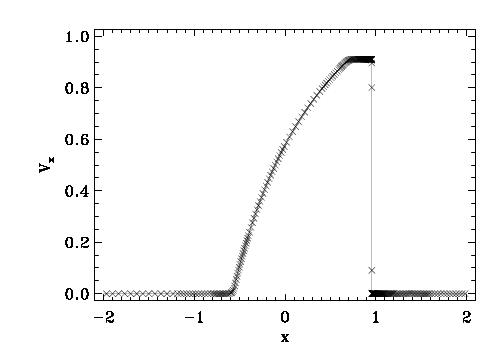}}}}
{{\resizebox{0.48\textwidth}{!}{\includegraphics{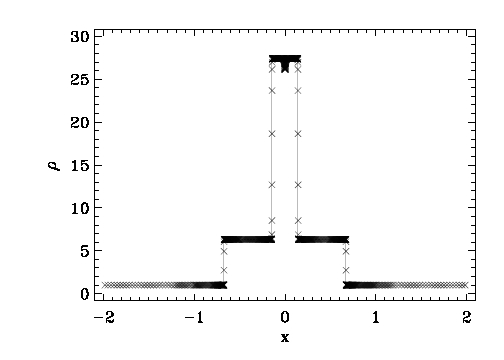}}}}
{{\resizebox{0.48\textwidth}{!}{\includegraphics{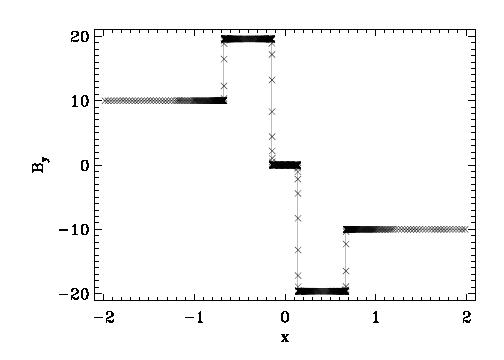}}}}
{{\resizebox{0.48\textwidth}{!}{\includegraphics{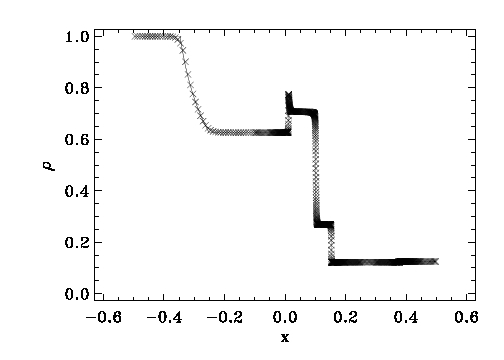}}}}
{{\resizebox{0.48\textwidth}{!}{\includegraphics{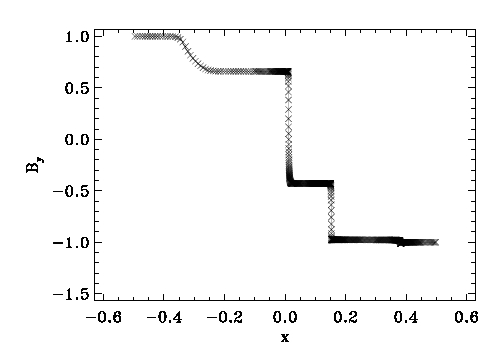}}}}
\end{center}
}
\caption{RMHD Riemann problems. Crosses indicate the AMR solutions, while solid lines are the exact solutions.}
\label{f-rp1}
\end{figure}

\begin{figure}
\FIG{
\begin{center}
{{\resizebox{0.48\textwidth}{!}{\includegraphics{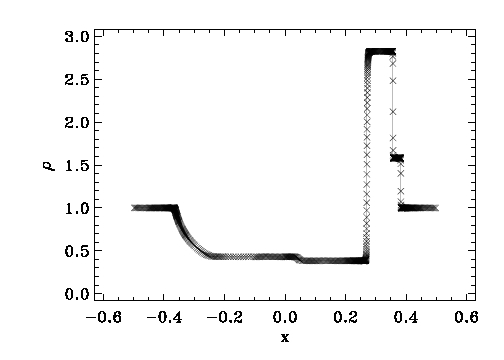}}}}
{{\resizebox{0.48\textwidth}{!}{\includegraphics{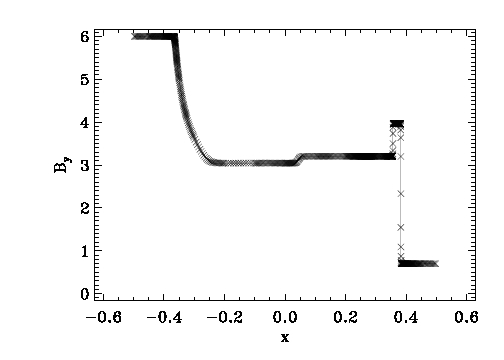}}}}
{{\resizebox{0.48\textwidth}{!}{\includegraphics{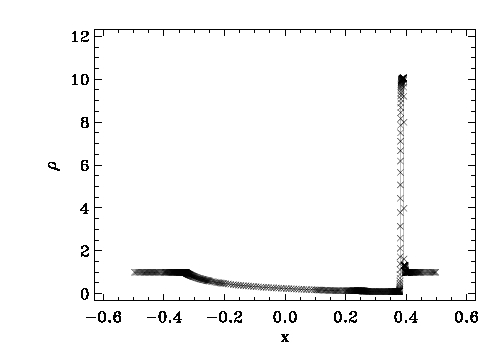}}}}
{{\resizebox{0.48\textwidth}{!}{\includegraphics{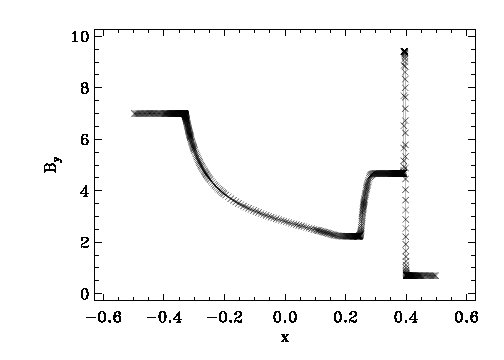}}}}
{{\resizebox{0.48\textwidth}{!}{\includegraphics{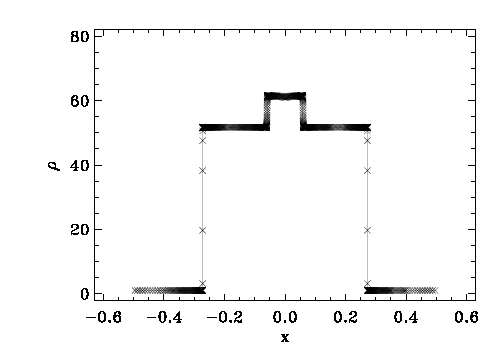}}}}
{{\resizebox{0.48\textwidth}{!}{\includegraphics{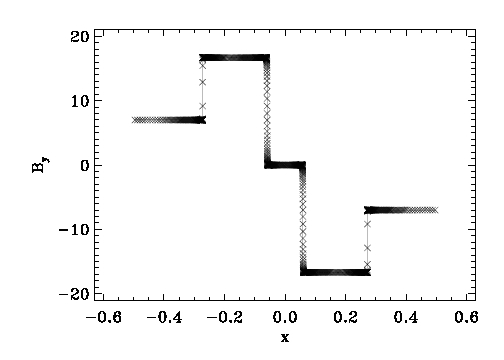}}}}
{{\resizebox{0.48\textwidth}{!}{\includegraphics{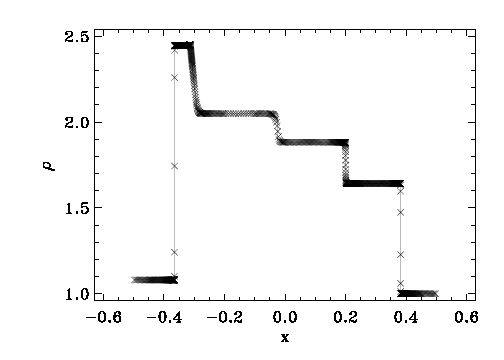}}}}
{{\resizebox{0.48\textwidth}{!}{\includegraphics{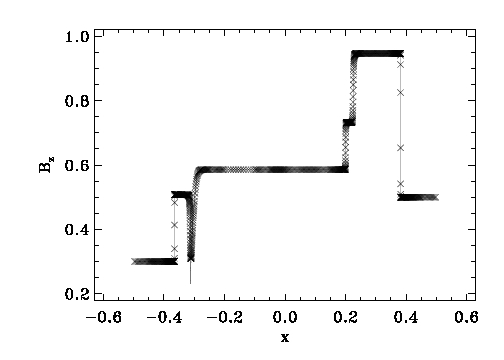}}}}
{{\resizebox{0.48\textwidth}{!}{\includegraphics{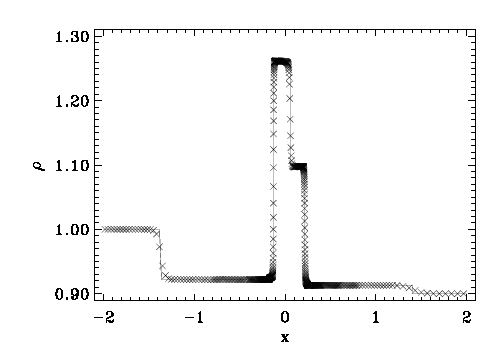}}}}
{{\resizebox{0.48\textwidth}{!}{\includegraphics{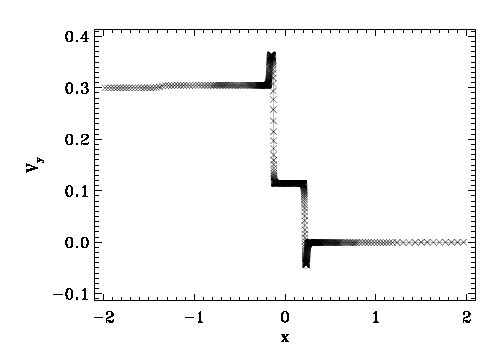}}}}
\end{center}
}
\caption{RMHD Riemann problems.}
\label{f-rp2}
\end{figure}

A direct validation of the code is provided by performing a series of Riemann
problem tests, where an initial discontinuity is left to evolve dynamically.
For relativistic MHD, a recent contribution by~\cite{giacom} documented how one
can obtain the exact `analytic' solution when accounting for the up to 7 wave
signals that may emerge out of the $t=0$ problem. The method was demonstrated
for 10 specific initial conditions, and we reproduce all these
10 cases numerically here. They collect various tests reported in recent
code developments for relativistic MHD \cite{komis,balsara}, augmented with some new Riemann problems.
Our results are shown in Figs.~\ref{f-rp1}-\ref{f-rp2}, where we overplot in all cases the exact solution generated
by~\cite{giacom}. We briefly comment on each case in what follows. We typically
compute with up to 8 grid levels, and use a base resolution of 60 unless stated otherwise. The exact initial
conditions are fully specified in~\cite{giacom}, and we refer to that paper
for details.

The first two tests have $B_x=0$ throughout, so that only two fast signals and a
tangential discontinuity emerges. The first two rows of Fig.~\ref{f-rp1} correspond
to the Shock-Tube 2, and Generic shock tube test, respectively. The first has a
left-going fast rarefaction, a tangential discontinuity, and a fast shock, and presents no major difficulty. The generic shock tube test has a leftgoing fast shock and rightgoing rarefaction,
with a tangential discontinuity in between. We find overshoot errors at both the
shock and the tangential discontinuity, which diminish only by raising the overall
resolution significantly: in Fig.~\ref{f-rp1} this test here exploited 240 base level
grid points, and we still have an erroneous variation in between shock and
tangential signal in $v_z$. 

The next set of 8 tests considers cases where $B_x\ne 0$. The third to fifth
row in Fig.~\ref{f-rp1} shows the outcome for the 3 coplanar problems, where
at most 5 wave signals can emerge. The third row has no $y-$ or $z-$velocity
nor magnetic field components, and represents a typical challenge when
a contact discontinuity is in close vicinity to a fast shock. Our AMR computation
with the TVDLF scheme adequately resolves the narrow structure, with a rather
unavoidable large number of grid points to represent the contact. The fourth test
is a collision leading to two pairs of left and rightgoing fast and slow shocks. 
Similar to all documented numerical solutions, we can hardly avoid to generate a
central error in density $\rho$, which should remain constant in between the two
slow shocks. The final case shown in Fig.~\ref{f-rp1} is the relativistic
analogue of the Brio-Wu test~\cite{briowu}. A left going rarefaction, a slow compound, a contact discontinuity, and a slow and fast shock are encountered from left to right. The correspondence with the analytic solution is satisfactory,
with many grid points representing the contact jump. Note
that the method to generate the `exact solution' excludes the possibility of compound waves, in part invalidating the comparison there.

The tests collected in Fig.~\ref{f-rp2} reproduce tests from~\cite{balsara} and
the generic Alfv\'en test introduced by \cite{giacom}. The top row has left
going fast and slow rarefactions, a contact and rightgoing slow and fast shocks.
Only the many grid points in the contact is arguable, but at the same time the
variation is captured at the highest grid level activated. The next test has a similarly structured outcome, with an even more extreme length contraction effect at play
between right-going contact, slow and fast shock. The latter are accurately 
captured at the correct amplitude, as best seen in the $B_y$ plot. The third row of
Fig.~\ref{f-rp2} yields another stringent collision test analogous to the one
shown in Fig.~\ref{f-rp1}. With no major differences in solution strategy, the
numerical result here (with base resolution 120) is far less polluted by a central density error.
The last two tests trigger all 7 wave signals from the initial discontinuity,
with Alfv\'en discontinuities in between the slow and fast signal pairs. At the times shown, the spacing between Alfv\'enic and slow signals can be very close
still. For the fourth row in Fig.~\ref{f-rp2}, the Balsara test 5 case, it is expected to be down to order $0.001$ for the leftgoing Alfv\'en discontinuity and slow rarefaction. It is seen in the $B_z$ plot how even higher effective resolution
would need to be used to capture this transition exactly. Still, we correctly
find all wave signals. The last generic Alfv\'en test has a similar challenge
for the Alfv\'en signals adjacent to slow shocks. The outmost
left-going fast rarefaction and right-going fast shock are captured on the lowest
grid resolution only for this test, which can be changed by fine-tuning
the employed refining criterion. In summary, our grid-adaptive numerical
solutions show a favorable agreement with the analytic results in all cases.

\subsection{Multidimensional tests}

In a first 2D test, we recompute the relativistic rotor problem, a test first
performed in non-relativistic MHD settings, and subsequently modeled in a relativistic variant by~\cite{delzanna}. We simulate the very same problem on a larger domain
$[0,2]^2$, in order to follow the rotor evolution at higher effective resolution to a longer time than presented originally. The
polytropic index is constant at $\gamma=5/3$, while initial pressure is unity throughout. 
A high density circular disk with $\rho=10$ rotates anti-clockwise at uniform angular velocity $\omega=9.95$ within radius $0.1$ from the center of the domain. The disk
discontinuously connects to a tenfold lower density, static medium. The entire
domain is pervaded by a homogeneous horizontal field ${\bf B}=\hat{e}_x$.
Using 7 refinement levels we acquire an effective resolution of $6400\times 6400$. This represents locally a four times higher resolution than used in~\cite{delzanna}, which compensates for the difference in order of accuracy
employed (a third order method versus our second order scheme, on smooth solutions). 

\begin{figure}
\FIG{
\begin{center}
{\resizebox{\textwidth}{!}{\includegraphics{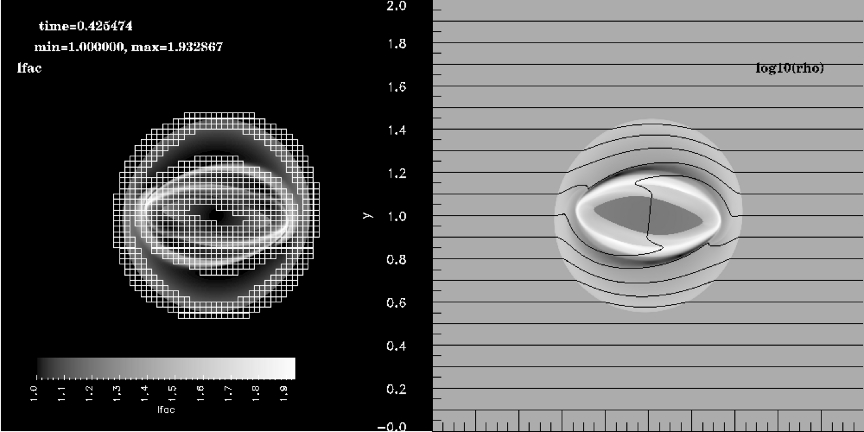}}}
{\resizebox{\textwidth}{!}{\includegraphics{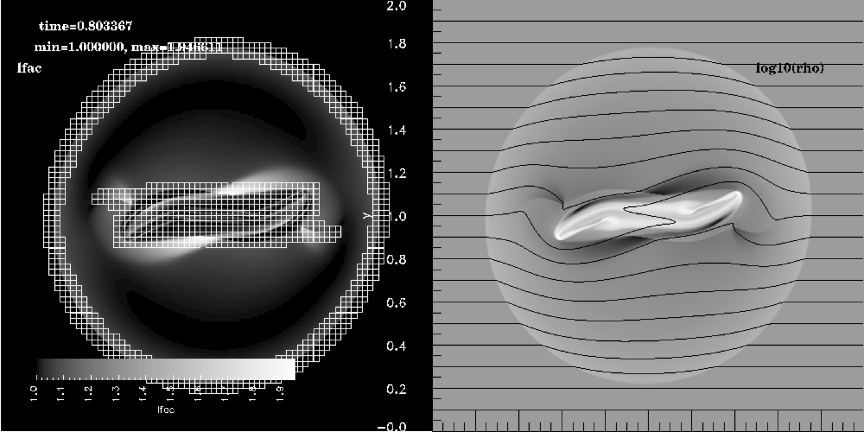}}}
\end{center}
}
\caption{The relativistic rotor at times $t=0.4$ and $t=0.8$, showing at left the Lorentz factor and the location of the intermediate level 5 grid blocks, and
the magnetic field lines on top of proper density distribution at right.}
\label{f-rot}
\end{figure}

We intentionally followed the rotor evolution to twice the time reported
earlier, till $t=0.8$. In~\cite{delzanna}, slight corrugations in density
could be detected cospatial with shear flow regions at time $t=0.4$, and we
intended to investigate their potential role in any further nonlinear evolution.
Snapshots at $t=0.4$ and $t=0.8$ are shown in Fig.~\ref{f-rot}. Fast shock fronts can be detected to travel outwards into the static surroundings, and inwards
towards the disk center. Slow rarefactions are found in between, and the field
deflections in effect brake the initial fast rotation ($\Gamma\simeq 10$). 
The overall evolution remains pointsymmetric. At $t=0.4$ the field in the disk has rotated over about $90^\circ$, and the in- and outwards traveling shocks
can be clearly detected. We found somewhat higher instantaneous Lorentz factors
than those reported before, and no evidence of density corrugations. Also,
as shown in the second snapshot where the inward shocks have already collided,
no clear indication of a shear-induced fine structure was found. 
The automated grid refinement does follow the density variations at the highest
grid resolution throughout the computation. An impression of the location of
the intermediate level 5 grids (two more grid levels exist on top of this
level) is shown in Fig.~\ref{f-rot}. 

\begin{figure}
\FIG{
\begin{center}
{\resizebox{0.45\textwidth}{5cm}{\includegraphics{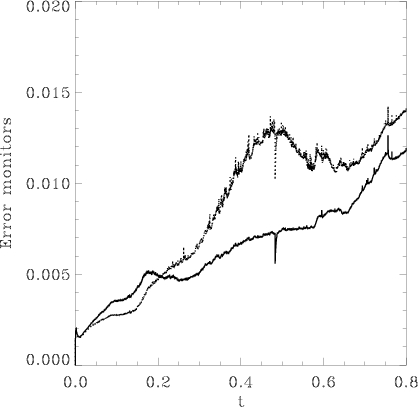}}}
{\resizebox{0.48\textwidth}{6cm}{\includegraphics{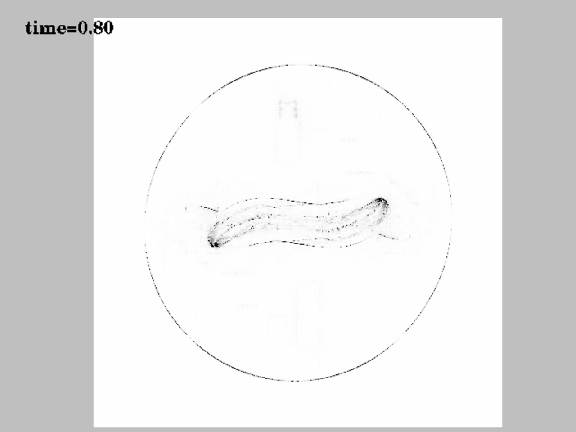}}}
\end{center}
}
\caption{The stabilizing role of the parabolic monopole treatment illustrated. At left, we show the temporal evolution of the monopole error monitors
$\frac{1}{V}\int | \nabla \cdot \bfB | \,dx dy$ (solid line) and $\frac{1}{V}\int (| \nabla \cdot \bfB |)/(\parallel \bfB \parallel) \,dx dy$ (dotted line). At right, the instantaneous distribution of the monopole errors at the final time of the simulation. In all these monitors, the divergence is evaluated discretely as a centered difference formula.}
\label{f-rotdivb}
\end{figure}

To illustrate the magnetic monopole control used in this particular simulation, we provide various quantitative measures
for it in Fig.~\ref{f-rotdivb}. As stated earlier, the parabolic treatment is meant to stabilize the computation and uses
a discrete evaluation of the magnetic field divergence in a diffusive type source term. The errors themselves are unavoidably
created continuously at the location of the strongest discontinuities. In this simulation, even the first timestep introduces
finite monopole errors at the border of the `rotating disk', since we effectively break the field lines there (e.g. the central
field line gets disconnected in two locations, where from one grid point to the next we jump from static to
Lorentz factor 10 flow). These monopoles can locally and temporarily be of order unity, while the diffusive treatment then
spreads and diffuses these errors during the computation. In Fig.~\ref{f-rotdivb}, the left panel shows two domain 
averaged error monitors as a function of time: it can be seen that these mean values remain at ${\cal O}(0.01)$ in this computation. The fact that they do not grow without bound is the most important observation, confirming their stabilizng role.
The right panel from Fig.~\ref{f-rotdivb} is at the final time $t\approx 0.8$, and shows the instantaneous distribution of the monopoles, as evaluated in centered difference approximation.
Black values are locally order unity, and the largest errors necessarily coincide with the various shock wave fronts. Note that
our restriction and prolongation strategies do not particularly enforce solenoidal fields in any discrete sense, so
grid level boundaries can temporarily be detected in such error maps. Again, the role of the parabolic term then acts
to diffuse such errors at their maximal rate and it is important to note that this does not affect
any of the employed conservative variables adversely.

In a second 2D application, we generalize a shock-dominated time-dependent
problem frequently used in benchmarking classical MHD codes to the relativistic
regime. The non-relativistic test~\cite{piconedahl} considers a Mach 1 vortex superposed
on a multiple magnetic island configuration, on a doubly periodic $[0,2\pi]^2$ Cartesian domain. It starts from uniform density and pressure throughout, and the supersonic rotation concentrates magnetic field gradients in thin, localized
current sheets from which shock fronts originate, which subsequently interact.
Our relativistic analogue considers a relativistically hot gas, where the
internal energy dominates over the rest mass contribution, such that the relativistic sound speed approaches its maximal value $c_s\approx\sqrt{\gamma-1}$.
This is consistent with a polytropic index value $\gamma=4/3$, and we set the
initial pressure $p=10$ while proper density $\rho=1$. The vortex imposes a
velocity field
\begin{equation} {\bf v}=-A \sin{(y)} \hat{e}_x + A \sin{(x)} \hat{e}_y,
\end{equation}
where $A=0.99/\sqrt{2}$, ensuring subluminal and supersonic velocities. The maximal initial
Lorentz factor is then about 7. The magnetic field is then initialized at
\begin{equation}
{\bf B}=- \sin{(y)} \hat{e}_x +  \sin{(2x)} \hat{e}_y,
\end{equation}
which makes the ratio of magnetic to thermal pressure attain a maximal value of
$\beta_I=0.098$. We simulated this problem for times beyond $t=12.5$, at which time the maximal Lorentz factor encountered has dropped to about $\Gamma=1.526$. 
The simulation used a $40\times 40$ base resolution, with a maximal 7 refinement
levels, effectively mimicking a $2560\times 2560$ resolution.

\begin{figure}
\FIG{
\begin{center}
{\resizebox{0.49\textwidth}{!}{\includegraphics{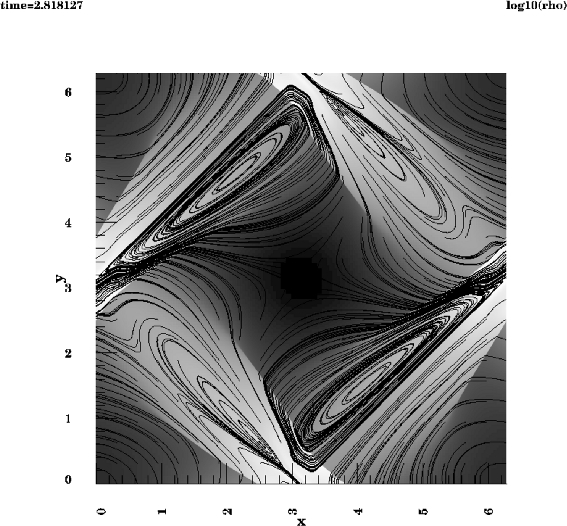}}}
{\resizebox{0.49\textwidth}{!}{\includegraphics{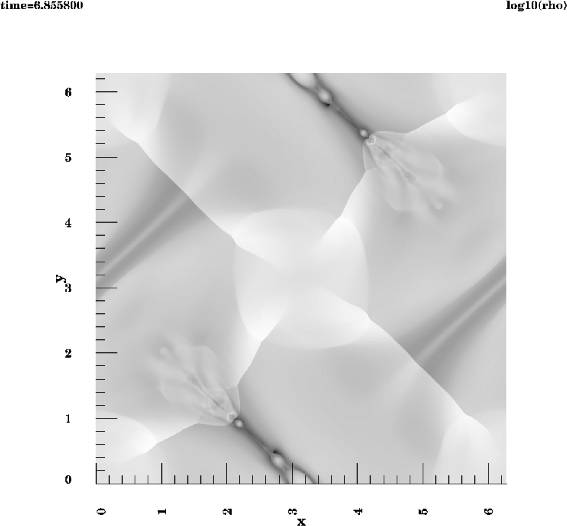}}}
{\resizebox{0.49\textwidth}{!}{\includegraphics{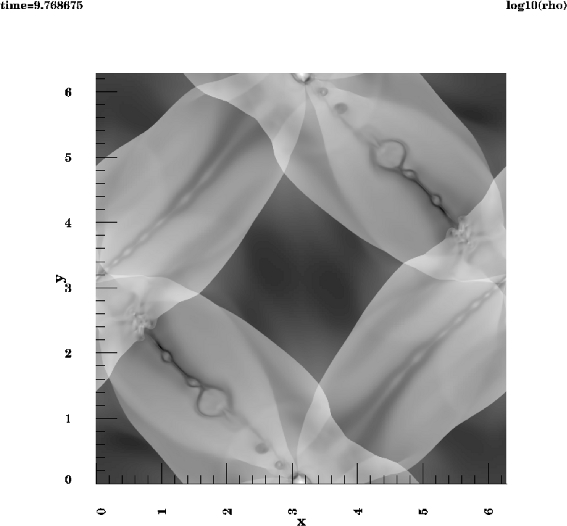}}}
{\resizebox{0.49\textwidth}{!}{\includegraphics{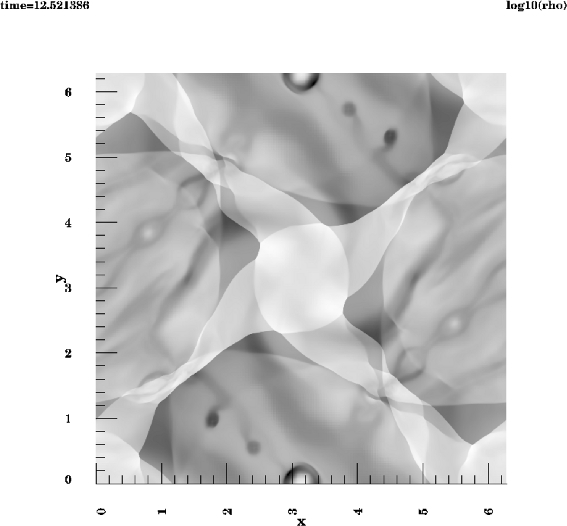}}}
\end{center}
}
\caption{The shock-dominated distortion of a magnetic island chain, in a relativistic, supersonic analogue of the compressible variant of the Orszag-Tang vortex. 
Shown at consecutive times is the logarithm of the proper density, with an impression of the island deformation in the first panel. See text for details.}
\label{f-ot}
\end{figure}

The initial magnetic topology is characterized by alternating $X$ and $O$ nul-points (where the field vanishes), with 4 different $X$ and 4 $O$ type nuls in the doubly periodic domain.
Along the $y=\pi$ horizontal, we thus encounter two islands of closed field
(to the left and right of the central $X$ point), and this
pattern repeats on $y=0,2\pi$ with distinct $\pi/2$ phase difference.
The superposed vortical velocity will immediately displace the left-central 
island in a diagonally upwards fashion, while distorting the right-central
island diagonally downwards. The double periodicity implies that the island
situated midway the horizontal boundaries gets squashed in the process,
and oriented in a diagonally downwards current sheet. This violent compression 
will drive two shock waves, one from either side of this sheet, traveling against the original flow direction. In a similar fashion, the flow induces
a strong shearing at the $X$ point midway the vertical sides, with similar
accumulation of matter along a diagonally upward pointing sheet. There too,
two shock fronts separate off the compression zone. The first panel of
Fig.~\ref{f-ot} at $t=2.82$ superposes the field structure on the (logarithmic) proper density, clearly showing the four diagonal shock fronts and the island deformations
just described.  These four interacting shock fronts meet up in the centre of the domain, while the sheet formed in the compressed $O$ configuration eventually
demonstrates a spontaneous break-up forming a series of islands. This tearing type reconnection happens at about $t=4.6$. In this sudden topological magnetic reconfiguration, some of the smaller islands get accelerated towards the shock
fronts traveling away from the sheared and compressed $X$ point. In the second
panel of Fig.~\ref{f-ot} at $t=6.85$, some of these islands and the shock front
deformations they cause can be seen. These and similar interactions occuring
with the diagonal shock fronts converging to the sheared $X$ sheet, drive a
second sequence of strong density variations colliding upon both the compressed
$O$ point (now with its island structure) and the sheared $X$ sheet. A second quadruplet of shocks then propagates away from these locations. This causes intricate shock-shock interference patterns with the original 4 shock fronts.
The sheared $X$ sheet in fact also gets torn up into an island chain structure,
as seen in the third panel from Fig.~\ref{f-ot}. Eventually, also this second
shock sequence meets up at the centre, while only some of the islands from the original $O$ and $X$ sheet tearing events survive as localized
density enhancements or depletions. The final panel of Fig.~\ref{f-ot} shows
the by now rather complicated density distribution near time $t=12.5$.
The point-symmetry about the centre of the domain is preserved perfectly throughout the entire grid-adaptive computation.
Note that this long-term computation gives strong evidence for intricate reconnection events,
which will be influenced by the numerical scheme employed: it could be of interest to benchmark several higher order schemes, in combination
with varying strategies for magnetic monopole control, on this problem in particular. 
Ultimately, deviations from perfect conductivity would need to be explicitly accounted for.

\begin{figure}
\FIG{
\begin{center}
{\resizebox{\textwidth}{5cm}{\includegraphics{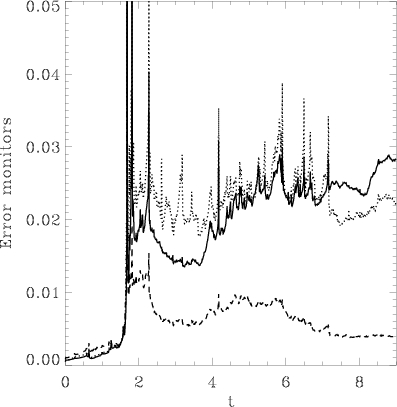}}}
\end{center}
}
\caption{The stabilizing role of the parabolic monopole treatment illustrated for the second test case. We quantify the temporal evolution of three monopole error monitors
$\frac{1}{V}\int | \nabla \cdot \bfB | \,dx dy$ (solid line), $\frac{1}{V}\int (| \nabla \cdot \bfB |)/(\parallel \bfB \parallel) \,dx dy$ (dotted line), and $\frac{1}{V}\int (| \nabla \cdot \bfB |)/(\parallel \nabla \parallel \bfB \parallel\parallel) \,dx dy$ (dashed line). In all these monitors, the divergence is evaluated discretely as a centered difference formula.}
\label{f-otdivb}
\end{figure}

As a final illustration of the parabolic monopole treatment used in this test, Fig.~\ref{f-otdivb} collects the temporal evolution of various domain averaged error monitors 
over most of the simulated period. Once more, the error diffusion approach
works as intended, despite the presence of very strong interacting shock fronts, the sudden appearance of magnetic island 
chains, and the continuously adjusting AMR grid hierarchy.


\section{Conclusion}\label{sec-summ}
We provided details on our relativistic grid-adaptive MHD code, and tested it using recently
available Riemann problem solutions, as well as multi-dimensional setups. The latter include
a new long-term simulation demonstrating shock-shock interactions and reconnection events, and
this could be of interest to benchmark existing shock-capturing algorithms on long-term
shock-dominated relativistic magnetized flow problems, such as those recently presented in~\cite{komis2}.

In future work, the RMHD code AMRVAC will be used to model AGN jet propagation~\cite{jetsrmhd} and
GRB dynamics. The magnetic field
is suggested as possible ingredient to achieve the high Lorentz factor flows, such as reached by
the GRB fireball, and plays a crucial role in both AGN and GRB flow collimation. 
Other future projects concentrate on
implementing a more realistic equation of state to 
investigate the launching and
the propagation of relativistic jet. We will discuss these 
astrophysical applications with more detail in future publications.

\section*{Acknowledgments}

We acknowledge the use of an exact Relativistic Riemann solver from Bruno Giacomazzo, for overplotting the exact solutions in Fig.~\ref{f-rp1}-\ref{f-rp2}.
Computations have been performed on the K.U.Leuven High Performance Computing cluster VIC. 
ZM and RK acknowledge financial support from the Netherlands Organization for Scientific
Research, NWO grant 614.000.421, and `Stichting voor Fundamenteel Onderzoek der Materie' FOM.

\end{document}